\documentclass[aps,prl,floatfix,twocolumn]{revtex4}
\usepackage{amsmath}
\usepackage{epsfig}
\usepackage{subfigure}
\usepackage{graphicx}
\usepackage{units}
\usepackage{epstopdf}

\long\def\symbolfootnote[#1]#2{\begingroup%
\def\thefootnote{\fnsymbol{footnote}}\footnote[#1]{#2}\endgroup}
\DeclareMathOperator{\Tr}{Tr}
\DeclareMathOperator{\Real}{Re}
\DeclareMathOperator{\Imag}{Im}

\begin{document}

\title{Experimental quantum computing without entanglement}

\author{B. P. Lanyon, M. Barbieri, M. P. Almeida and A. G. White}
\affiliation{Department of Physics and Centre for Quantum Computer Technology, University of Queensland, Brisbane 4072, Australia}

\begin{abstract}
\noindent Entanglement is widely believed to lie at the heart of the advantages offered by a quantum computer. This belief is supported by the discovery that a noiseless (pure) state quantum computer must generate a large amount of entanglement in order to offer any speed up over a classical computer. However, deterministic quantum computation with one pure qubit (DQC1), which employs noisy (mixed) states, is an efficient model  that generates at most a marginal amount of entanglement. Although this model cannot implement any arbitrary algorithm it can efficiently solve a range of problems of significant importance to the scientific community. Here we experimentally implement a first-order case of a key DQC1 algorithm and explicitly characterise the non-classical correlations generated. Our results show that while there is no entanglement the algorithm does give rise to other non-classical correlations, which we quantify using the quantum discord---a stronger measure of non-classical correlations that includes entanglement as a subset. Our results suggest that discord could replace entanglement as a necessary resource for a quantum computational speed-up. Furthermore, DQC1 is far less resource intensive than universal quantum computing and our implementation in a scalable architecture highlights the model as a practical short-term goal.
\end{abstract}

\maketitle

In contrast to the highly pure multi-qubit states required for the conventional models of quantum computing \cite{MikeIke, PhysRevLett.86.5188}, DQC1 employs only a single qubit in a pure state, alongside a register of qubits in the fully mixed state \cite{PhysRevLett.81.5672}. While this model is strictly less powerful than a universal quantum computer  (where one can implement any arbitrary algorithm) it can still efficiently solve important problems that are thought to be classically intractable. The application originally identified was the efficient simulation of some quantum systems \cite{PhysRevLett.81.5672}. Since then \textit{exponential speed-ups} have been identified in estimating: the average fidelity decay under quantum maps \cite{Poulin2004a}; 
quadratically signed weight enumerators \cite{Knill:2001fp}; and the Jones Polynomial in knot theory \cite{shor-2007}. Recently it has been shown that DQC1 also affords efficient parameter estimation at the quantum metrology limit \cite{somma-2007}. Furthermore, attempts to find an efficient way of classically simulating DQC1 have 
failed \cite{datta:042310}. These results provide strong evidence that a device capable of implenting scalable DQC1 algorithms would be an extremely useful tool.

Besides the practical applications, DQC1 is also fascinating from a fundamental perspective. For example, it is straightforward to show that a model employing only fully mixed qubits offers no advantage over a classical computer. It is therefore surprising that the addition of only a single pure qubit offers such a dramatic increase in computational power. 
Furthermore, an important quantum information result is that a pure state quantum computer can only offer an advantage over a classical approach if it generates an amount of entanglement that grows with the size of the problem being tackled \cite{joszalinden, PhysRevLett.91.147902}. This supports the commonly held belief that entanglement lies at the heart of the quantum advantage. However entanglement is at most marginally present in DQC1 \cite{datta:042316}.  The existence of efficient DQC1 algorithms therefore provides strong evidence that a large amount of entanglement is not necessary to achieve a speed-up. Instead it has been proposed that a requirement, common to both pure and mixed state models, is the generation of quantum correlations that are not fully captured by entanglement \cite{datta:050502, datta:042310}. It is now important to pursue a deeper understanding of these correlations.

\begin{figure}
\begin{center}
\includegraphics[width=0.5 \columnwidth]{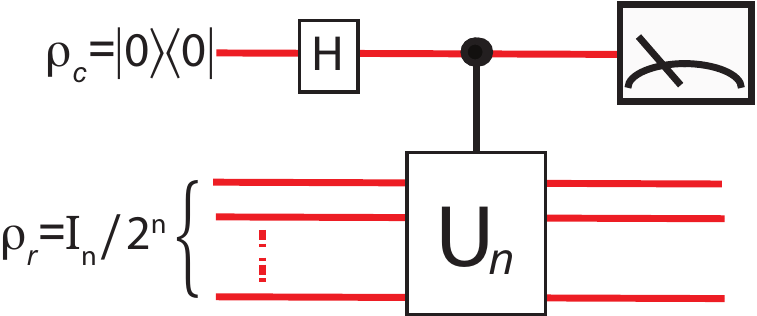}
\end{center}
\vspace{-3mm}
\caption{DQC1 normalised trace estimation algorithm. Repeated running of the circuit and measurement of qubit $c$ yields estimates of the standard Pauli $\langle X \rangle$ and $\langle Y \rangle$ expectation values, from which the normalised trace of the unitary operator $U_n$ can be derived ($\Tr[U_n]/2^n$). $I_n$ is the n qubit identity.
}
\vspace{-3mm}
\label{fig:bell}
\end{figure}

Following recent developments in experimental quantum computing \cite{toffolipaper} we  present an implementation of DQC1 in a linear optic architecture. We build on a previous demonstration in liquid state NMR \cite{ryan:250502} in two important ways: we are able to characterise the quantum correlations generated by our implementation; there are several known paths to efficient scalable linear optic quantum computing \cite{Knill:2001lr, PhysRevLett.86.5188, kok:135} and there is much recent and ongoing progress towards developing the necessary technology \cite{grangier, dowling, AlbertoPoliti05022008}.

We present a first-order implementation of the DQC1 algorithm for estimating the normalised trace of a unitary matrix \cite{PhysRevLett.81.5672, datta:042316, datta:050502, datta:042310}. This algorithm achieves an exponential speedup over the best known classical approach i.e. it requires exponentially fewer resources as the size of the unitary increases. It is thought highly unlikely that an efficient, but as yet unknown, classical approach can exist \cite{datta:042316}. That DQC1 can perform this task efficiently underpins the model's ability to solve the aforementioned range of important practical problems.  

Figure~1 shows the normalised trace estimation algorithm. The required input state is separable and consists of a single pure qubit $c$ in the logical state $|{0}\rangle\langle{0}|{=} \frac{1}{2}\{I_1+Z\}$, and a register of $n$ qubits in the completely mixed state $I_n/2^n$. The circuit consists of the standard Hadamard gate \cite{MikeIke} and a unitary ($U_n$) on the register, controlled by qubit $c$. The state of all $n{+}1$ qubits at the output of the circuit is:
\begin{equation}
\rho_{cr}=
\frac{1}{2N}
\left[
\begin{matrix}
 I_n & U_{n}^\dagger  \\
   U_{n} &  I_n 
\end{matrix}
\right]
\end{equation}
where $N{=}2^n$ and $I_n$ is the $n$ qubit identity. The reduced state of qubit $c$ (after performing a partial trace over the register) is given by:

\begin{equation}
\rho_c=
\frac{1}{2}
\left[
\begin{matrix}
   1 & \Tr[U_n]^\dagger /N \\
   \Tr[U_n] /N&  1 
\end{matrix}
\right]
\end{equation}

\noindent Thus the normalised trace of $U_n$ is encoded in the coherences of qubit $c$. This information can be retrieved by measuring the expectation values of the standard Pauli operators X and  Y, since $\langle X \rangle{=}\Real[Tr(U_n)/2N]$ and $\langle Y \rangle{=}-\Imag[Tr(U_n)/2N]$. 

An expectation value is estimated by repeatedly running the circuit. One can achieve a fixed accuracy $\epsilon$ in this estimate with a number of runs $L\sim \ln(1/P_e)/\epsilon^2$, where $P_e$ is the probability that the estimate is farther from the true value than $\epsilon$ (Ref.~ \cite{datta:042316}). That the accuracy does \textit{not} scale with the size of the unitary and scales logarithmically with the error probability means that this is an efficient algorithm for estimating the normalised trace. In contrast, classical approaches suffer an exponential increase in required number of resources with the size of the 
unitary \cite{datta:042316}. 

\begin{figure}
\begin{center}
\includegraphics[width=1 \columnwidth]{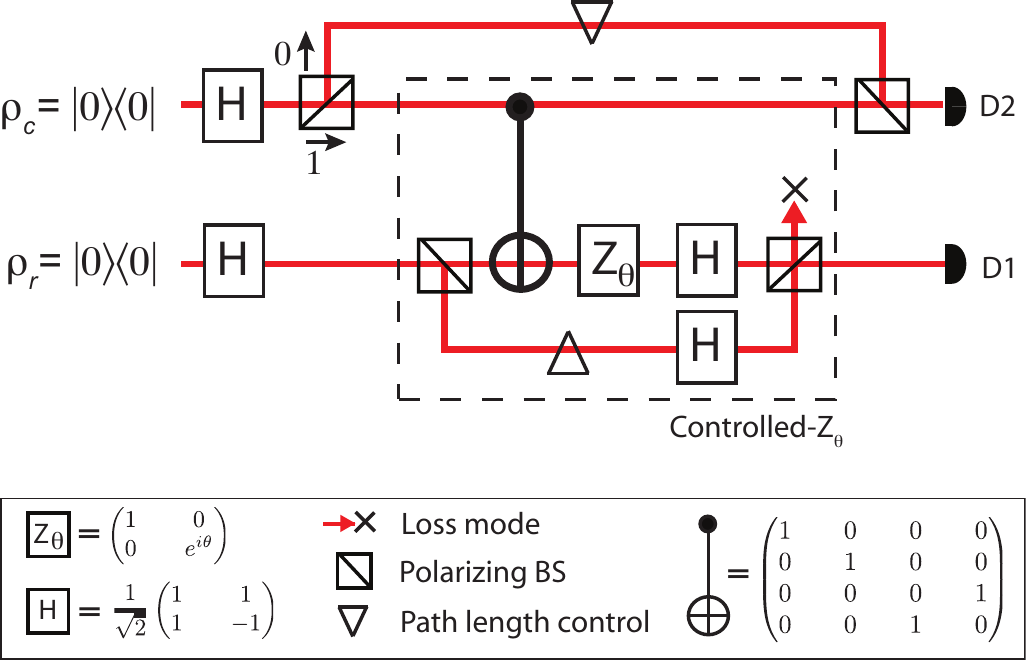}
\end{center}
\vspace{-3mm}
\caption{
Experimental schematic. The controlled-$\textsc{z}_{\theta}$ gate is implemented using a recently developed technique \cite{toffolipaper}. At the input and output qubits are encoded in the polarisation of single photons ($|{0}\rangle{=}|{\textsc{h}}\rangle,|{1}\rangle{=}|{\textsc{v}}\rangle$, Horizontal and Vertical). Coincident measurement of single photons at detectors D1-2 signals a successful run of the gate. The interferometers in each mode allow the preparation of noisy (mixed) states, by introducing a path difference between the two arms. See Methods.}
\vspace{-3mm}
\label{fig:bell}
\end{figure}

We implement the first-order ($n{=}1$) case for the unitary:
\begin{equation}
U_1= Z_{\theta}=
\left[
\begin{matrix}
      1 & 0  \\
   0 & e^{i\theta} 
\end{matrix}
\right]
\end{equation}

\noindent In this case $\langle X \rangle{=}(1{+}\cos\theta)/2$ and $\langle Y \rangle{=}(\sin\theta)/2$. 

Our implementation is performed in an all optical architecture shown in Fig.~2 (see Methods). We encode quantum information in multiple degrees of freedom of single photons. Single qubit gates are realised deterministically using birefringent wave-plates. The required two-qubit gate is realised non-deterministically and measurement of single photons in the two output modes signals a successful run \cite{toffolipaper}.   

The trace estimation algorithm is implemented over the range $-\pi {\leq} \theta {\leq} \pi$ (Eqn.~3).  Fig.~3a compares the experimentally observed results with the ideal. The ideal case is calculated assuming perfect circuit operation and measured input states. We observe a high degree of correlation quantified by a reduced $\chi^2$ of 0.7 (real curve) and 1.2 (imaginary curve). Deviations from the ideal are due to imperfect circuit operation caused by: optical beam steering as $\theta$ is varied; interferometric instability; and non-classical interference instability. There are several known paths to reduce these errors including moving to micro-optic systems \cite{AlbertoPoliti05022008}.

Interestingly, the exponential speed-up offered by this algorithm is not affected by reducing the purity of qubit $c$ \cite{datta:042316}. Consider replacing the initial state of this qubit with the mixed state $\frac{1}{2}\{I_1+\alpha Z\}$, where $\alpha$ now reflects the purity ($p{=}[1+\alpha^2]/2, 0{\le} \alpha {\le}1$). At the output of the circuit the state of this qubit is now given by:
\begin{equation}
\rho_{c'}=
\left[
\begin{matrix}
   0.5 & \alpha Tr(U_n)^\dagger /N \\
   \alpha Tr(U_n)/N &  0.5 
\end{matrix}
\right]
\end{equation}
The effect of mixture in qubit $c$ is to reduce the expectation values $\langle X\rangle$ and $\langle Y\rangle$ by $\alpha$ (Eqn.~2), thereby making it harder to estimate the normalised trace. In order to achieve the same fixed accuracy as before requires an increased number of runs  $L'{\sim}L/\alpha^2$. While this clearly adds an additional overhead, as long as $\alpha$ is non-zero, the algorithm still provides an efficient evaluation of the trace. Even access to the tiniest fraction of a single pure qubit is sufficient to achieve an exponential speedup over the best known classical approach.

Fig.~3b compares the experimentally observed algorithm results with the ideal, for $\alpha{=}0.58{\pm}{0.02}$. We observe a high correlation with the ideal, quantified by a reduced $\chi^2$ of 1.8 (real curve) and 2.0 (imaginary curve). The increased $\chi^2$ in this case (compared to the results in Fig.~3a) is due to a less favorable optical alignment and is not an intrinsic error associated with initialising $c$ into a mixed state. The additional resource overhead in this case is apparent in the amplitude reduction by a factor of $\alpha$ compared with the results shown in Fig.~3a.

\begin{figure}
\begin{center}
\includegraphics[width=1 \columnwidth]{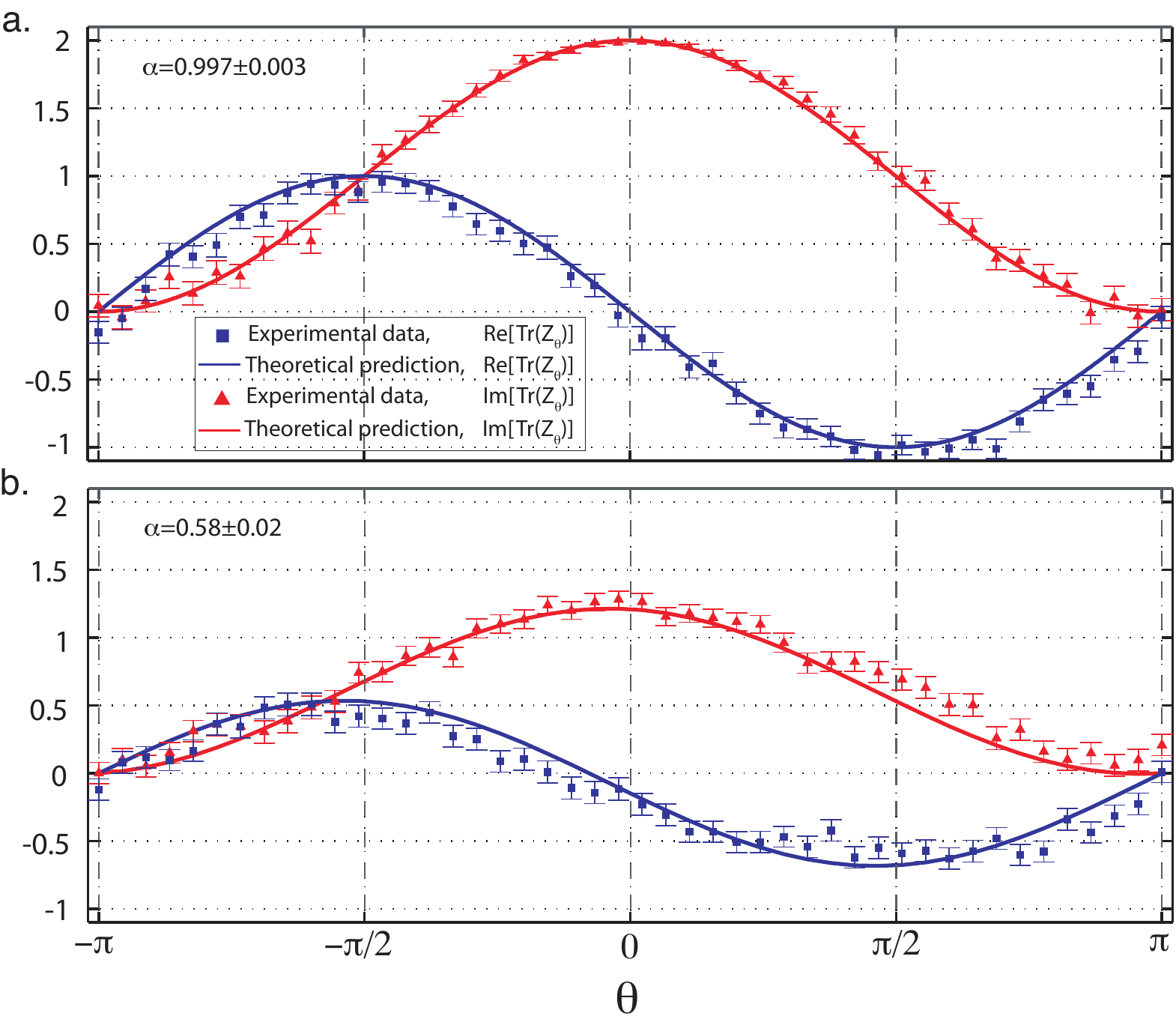}
\end{center}
\vspace{-5mm}
\caption{Algorithm output. Real and imaginary parts of the renormalised trace measured for two values of $\alpha$, over a range of $\theta$, Eqn.~4. Theoretical predictions are calculated using measured input states and assuming perfect circuit operation. We observe a $\chi^2$ of \textbf{a}, 0.95 and \textbf{b}, 2.0. $\langle X\rangle$ is estimated by counting the number of coincident photon pairs ($N$) when projecting qubit $c$ into the states $|{\pm{}}\rangle{=}(|0\rangle{\pm{}}|1\rangle)/\sqrt{2}$, over 10 secs. Then $\langle X\rangle{=}(N_+{-}N_-)/(N_+{+}N_-)$. The same technique is used to estimate $\langle Y\rangle$, but in this case we project into the states $|{\pm{i}}\rangle{=}(|0\rangle{\pm{}}i|1\rangle)/\sqrt{2}$. }
\vspace{-5mm}
\label{fig:bell}
\end{figure}

Finally we analyse the correlations generated by the algorithm by performing full tomography of the two-qubit output state before the final measurement stage (Eqn.~1), using 36 (over-complete) measurement bases. This allows a reconstruction of the density matrix. Figure~4 shows two measures of non-classical correlations, derived from the experimentally reconstructed density matrices---the well-known tangle \cite{PhysRevA.61.052306} and the lesser-known quantum discord \cite{0305-4470-34-35-315, PhysRevLett.88.017901, datta:050502}. The tangle is a complete measure of entanglement in two-qubit states and represents perhaps the most striking divergence from classical behavior. However, entanglement is not the only kind of non-classical correlation. A far stronger measure, that encompasses entanglement and more, is given by the quantum discord. 

The discord is concerned with a fundamental characteristic of classical systems---that their information content is locally accessible and can be obtained without perturbing the state for an independent observer \cite{PhysRevLett.88.017901}.  If the discord is zero there exists a local measurement protocol under which all the state information can be revealed without perturbing the state for observers who do not have access to the measurement results.  If the discord is non-zero then no such protocol exists. For pure states, discord becomes a measure of entanglement---no other non-classical correlations can be distinguished in this case. However, for mixed states the discord captures more non-classical correlations than entanglement \cite{datta:050502}. 

The results show that, to within experimental error, our implementation does not give rise to any entanglement. However, in general it does generate quantum discord. We observe a high degree of correlation between the theoretical and measured discord values, quantified by a $\chi^2$ of 1.6. 
These results are consistent with recent theoretical work \cite{datta:050502} which predicts that, although the entanglement is generally zero for arbitrary instances of this algorithm, discord is consistently present. 

For our implementation the discord is zero in two distinct cases $\theta{=}\{0,\pm{\pi}\}$, corresponding to our controlled-$\textsc{z}_{\theta}$ gate implementing the identity and a controlled-$\textsc{z}_{\pi}$, respectively. Both of these gates, alongside the Hadamard, are members of the Clifford group \cite{MikeIke}. 
In these cases the entire state evolution is implemented by gates from the Clifford group. Furthermore, the algorithm involves preparing the input in a mixture of logical basis states and measurement of observables in the Pauli group \cite{MikeIke}. Under these conditions the Gottesman-Knill theorem tells us that the entire algorithm can always be efficiently simulated on a classical computer \cite{MikeIke} (see supplementary material). In contrast, for all other values of $\theta$ the action of the controlled-$\textsc{z}_{\theta}$ gate is responsible for a non-Clifford group evolution. There is no known classical method to efficiently simulate an arbitrary size algorithm that evolves in this way---thereby allowing for a quantum speed-up. Furthermore, it is straightforward to show that any order implementation of the normalised trace estimation algorithm that is composed entirely of gates from the Clifford group produces a state with zero discord (see supplementary material). These results provide evidence of a link between discord and the potential for a DQC1 speedup.  An important path for further research is to determine whether all DQC1 circuits that do not generate discord can be efficiently simulated on a classical computer. Such a result would provide strong evidence that the discord is a more accurate measure of the resources required for a quantum speed up, than entanglement. Besides the fundamental interest, this could have implications in the many burgeoning quantum computing architectures where environmental decoherence presents a significant obstacle to universal pure state quantum computing.

We have generated, and fully characterised, quantum correlations that lie beyond entanglement and provided evidence that they enable efficient quantum computation. It is now of interest to explore quantum discord in the context of other active research areas, such as `non-locality without entanglement' \cite{PhysRevA.59.1070, pryde:220406}: while the two-qubit states of interest in these works are not entangled they all have non-zero discord, signifying the presence of quantum correlations. \\

\noindent The authors wish to thank C. Caves, A. Datta, A, Shaji, and G. Vidal for discussions. We acknowledge financial support from the Australian Research Council Discovery and Federation Fellow programs, the DEST Endeavour Europe and International Linkage programs, and an IARPA-funded U.S. Army Research Office Contract. Correspondence and requests for materials should be addressed to BPL (lanyon@physics.uq.edu.au). \\

\begin{figure}
\vspace{-4mm}
\begin{center}
\includegraphics[width=1 \columnwidth]{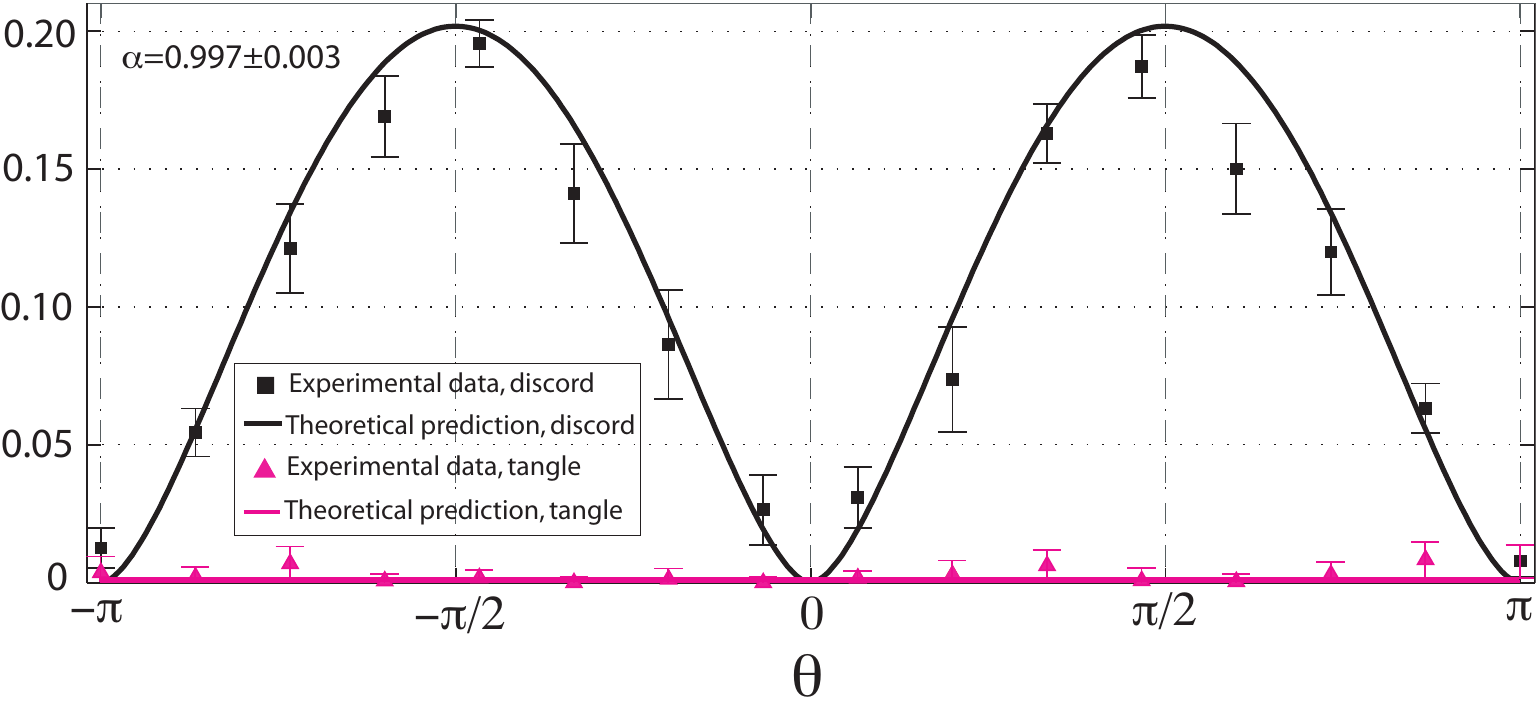}
\end{center}
\vspace{-5mm}
\caption{Non-classical correlations generated by the algorithm. Discord and tangle derived from experimentally reconstructed density matrices measured at the algorithm output for $\alpha{=}0.997{\pm}{0.003}$ (Fig.~3a). The discord is calculated by optimising over all 1-qubit local projective measurements on qubit c, Fig~1 (see supplementary material). Theoretical predictions are calculated using measured input states and assuming perfect circuit operation. The observed $\chi^2$ for the discord is 1.6. 
}
\vspace{-5mm}
\label{fig:bell}
\end{figure}

\small{\section{Methods}\
Photons are generated via spontaneous parametric down conversion of a frequency-doubled mode-locked Ti:Sapphire laser (820nm{$\rightarrow$}410nm, $\Delta \tau{=}80$fs at 82MHz) through a Type-I 2mm BiB$_{3}$O$_{6}$ crystal. Photons are filtered by blocked interference filters at 820$\pm{1.5}$nm; collected into two single-mode optical fibers; injected into free-space modes $c$ and $r$ (Fig.~2); detected using fibre-coupled single photon counting modules (D1--D2). With 100mW at 410nm we measure a two-fold coincidence rate at the output of the optical circuit of approximately $\unit[100]{s^{-1}}$. The required \textsc{cnot} is implemented using a standard technique involving non-classical interference at a partially polarising beamsplitter and projective measurement \cite{langford:210504, kiesel:210505, okamoto:210506}. State dependent loss is used to rebalance state amplitudes \cite{langford:210504, kiesel:210505, okamoto:210506}. 
To improve count rates, we achieve correct balance by pre-biasing the $c$ input state \cite{langford:210504, kiesel:210505, okamoto:210506}. The algorithm success probability is 1/12. We realise the required interferometers using calcite beam displacer pairs \cite{OBrien:2003lr} (Fig.~2).   

Each qubit goes through a polarisation interferometer (Fig.~2). We control the level of mixture in a qubit by altering the path-length difference between the arms of the corresponding interferometer. A path difference that is greater than the photon coherence length results in a fully decohered---that is, a fully mixed---photonic qubit.   By tuning the path difference between zero and the photon coherence length we can accurately control the level of mixture in the qubit between zero and maximum. We alter the path difference by rotating one calcite beam displacer of a pair about an axis perpendicular to the plane defined by the two paths.


All error bars are calculated through photon counting uncertainty described by poissonian statistics. We use the standard definition for a reduced $\chi^2$ calculation, allowing for three degrees of freedom (in our implementation both the real and imaginary parts of the trace are simple trigonometric functions defined by an amplitude, frequency and phase, Eqn.~3). \\

\noindent \section{Supplementary Information.}\
\\
\noindent \textbf{The Gottesman-Knill theorem for mixed states}.

\noindent The Gottesman-Knill theorem applies to algorithm input states prepared in the logical basis. In the DQC1 model the input is a mixture of logical basis states \cite{MikeIke}. However, it is straightforward to show that such an input state can be prepared from a pure logical state using only a number of additional Clifford group gates and ancilla qubits that is linear in the size of the input state. As such the DQC1 input state satisfies the conditions under which the theorem is valid.\\ 

\noindent \textbf{Definition of quantum discord for two-qubits}.

\noindent The definition of the mutual information for a bipartite density matrix is  \cite{datta:050502, MikeIke}:
\begin{equation}
{\cal I}(r{:}c){=}H(\rho_c)+H(\rho_r)-H(\rho_{cr}),
\end{equation}
where $H(\rho)$ is the well-known Von Neumann entropy of the state  \cite{MikeIke} $\rho$. An alternative definition is given by:
\begin{equation}
{\cal J}(r{:}c){=}H(\rho_r)-\tilde H(\rho_{cr}|c),
\end{equation}
where $\tilde H(\rho_{cr}|c)$ is the extension of the classical conditional entropy to the quantum case  \cite{datta:050502}. This is obtained by minimising the average entropy of the subsystem $r$, over all possible projective conditional measurements on $c$:
\begin{equation}
\tilde H(\rho_{cr}|c)=\min_{\{\Pi_i\}}\sum_i p_i H( \rho_{r|\Pi_i}),
\end{equation}
where $p_i{=}\text{Tr}(\Pi_i \rho_{cr} \Pi_i)$, and $\rho_{r|\Pi_i}= \text{Tr}_c(\Pi_i \rho_{cr} \Pi_i)/p_i$. The discord is defined as the difference  \cite{datta:050502}:
\begin{equation}
{\cal D}(r,c)={\cal I}(r{:}c){-}{\cal J}(r{:}c).
\end{equation}
Notice that ${\cal J}(r{:}c)$ is not symmetrical by inversion of $c$ and $r$, therefore, in general, discord is directional: ${\cal D}(r,c)\ne{\cal D}(c,r)$; we might not be able to detect quantum correlation when conditioning on measurements of one partition, while they arise  when considering the inverse case. It is straightforward to show that states admitting a diagonal representation in a local basis have bidirectionally vanishing discord---they contain only classical correlations. \\
\\
\noindent \textbf{Proof that DQC1 clifford evolution generates no discord}.

\noindent The DQC1 input state can be written in the form
\begin{equation}
\rho_\text{in}{=}\frac{1}{2^{n+1}}\left({\bf I}^{\otimes n+1}+{\bf Z}\otimes {\bf I}^{\otimes n}\right).
\end{equation}
which is clearly diagonal in the logic basis, hence it has zero discord in both directions. The action of Clifford group gates is to map Pauli matrices into Pauli matrices: if we indicate the unitary action of the circuit by $W$, the input state $\rho_{in}$ is transformed into:
\begin{equation}
\begin{aligned}
\label{out}
\rho_\text{cr}&{=}\frac{1}{2^{n+1}}\left({\bf I}^{\otimes n+1}+W({\bf Z} \otimes {\bf I} ^{\otimes n})W^\dag\right)\\
&{=}\frac{1}{2^{n+1}}\left({\bf I}^{\otimes n+1}+\bigotimes_{i=1}^{n+1}  \sigma_{r}^{(i)}\right),
\end{aligned}
\end{equation}
where ${\bf \sigma}_{r}^{(i)}$ refers to the $i$-th qubit, and r can take the values $r=0,1,2,3$ labeling the Pauli matrices:  $\sigma_0=\bf I$, $\sigma_1=\bf Z$, $\sigma_2=\bf X$, and $\sigma_3=\bf Y$. The state \eqref{out} is locally equivalent to the state:
\begin{equation}
\label{outish}
\rho'{=}\frac{1}{2^{n+1}}\left({\bf I}^{\otimes n+1}+\bigotimes_{i=1}^{n+1} \sigma_{s}^{(i)}\right),
\end{equation}
where the index $s$ can take only the values $s=0,1$; thus $\rho'$ admits a diagonalisation in a local basis. This state is purely classically correlated, hence its discord is zero.  Consequently $\rho_\text{out}$, which is obtained from $\rho'$ with local rotations, must have zero discord.
\\

\end{document}